\documentclass[journal]{IEEEtran}
\usepackage{cite}
\usepackage{amsmath,amssymb,amsfonts}
\usepackage{mathtools, cuted}
\usepackage{pifont}
\usepackage{algorithm}
\usepackage{algorithmic}
\usepackage{graphicx}
\usepackage{textcomp}
\usepackage{xcolor}
\usepackage[hidelinks]{hyperref}
\usepackage{orcidlink}
\usepackage{array}
\usepackage[caption=false,font=normalsize,labelfont=sf,textfont=sf]{subfig}
\usepackage{url}
\usepackage{verbatim}
\usepackage{booktabs}

\def\BibTeX{{\rm B\kern-.05em{\sc i\kern-.025em b}\kern-.08em
    T\kern-.1667em\lower.7ex\hbox{E}\kern-.125emX}}

\begin{document}

\title{Quantization-Induced Performance Degradation in Segment Routing:\\[-0.5ex] A Concentration Inequality Analysis}

\author{Kai Wang$^{\orcidlink{0000-0002-6170-4744}}$ and Chee Wei Tan$^{\orcidlink{0000-0002-6624-9752}}$
\thanks{Manuscript received xx November 2026; revised xx xxxx 202x; accepted xx xxxx 202x; approved by IEEE Communications Letter Associate Editor xxxx. xxxx. Date of publication xx xxxx 202x; date of current version xx xxxx 202x. This research has been supported in part by the Singapore Ministry of Education Academic Research Fund Tier 2 MOE-T2EP20224-0009 and NTU startup. \textit{(Corresponding author: Chee Wei Tan.)}}
\thanks{Kai Wang and Chee Wei Tan are associated with the College of Computing and Data Science (CCDS), Nanyang Technological University (NTU), Singapore 639798 (wang1899@e.ntu.edu.sg, cheewei.tan@ntu.edu.sg).}}

\markboth{IEEE COMMUNICATIONS LETTERS, VOL. 30, 2026}%
{Wang \MakeLowercase{\textit{et al.}}: Quantization-Induced Performance Degradation in Segment Routing: A Concentration Inequality Analysis}


\maketitle

\begin{abstract}
In segment routing (SR), splitting ratios must be quantized to respect practical hardware precision constraints. Yet no existing work has established an upper bound on the performance degradation caused by such quantization. In this work, we propose source-agnostic systematic rounding, which allows flows sharing a destination to share the same policy, thereby reducing memory consumption, and which approximates the target splitting ratios by a set of power-of-two terms so as to respect the hardware precision limit. We explicitly analyze the consequence of quantizing splitting ratios to integer multiples of $1/64$ by a probability-theoretic argument based on concentration inequalities, and we derive an upper bound on the resulting performance degradation. We validate our analysis with
extensive experiments on real-world network topologies and flow statistics.
\end{abstract}


\section{Introduction}\label{sec:introduction}

\IEEEPARstart{S}{egment} routing (SR) \cite{sr} is a source routing architecture that encodes an ordered segment list (SL) in the packet header to predefine the route at chosen granularity. The SR scheme allows only the ingress node to maintain SR policies and composes the SL, therefore simplifying traffic engineering (TE), improving network scalability and robustness, and reducing the burden on the control plane.

The SR architecture is specifically designed with the flexibility of the hardware infrastructure, either to leverage existing multi-protocol label switching (MPLS) network (SR-MPLS) with IGP-extensions and provide backward compatibility for a smoother migration, or to be built natively on an ubiquitous IPv6 network (SRv6) and unlock advanced network programming capabilities \cite{srsurvey}. There exist core differences between SR-MPLS and SRv6 in the data plane, for which SR-MPLS employs MPLS labels as segment identifiers (SIDs) and encodes the route as a label stack, whereas SRv6 utilizes 128-bit IPv6 addresses as SIDs and carries the SL through the segment routing header (SRH). Similar distinctions appear in the control plane, where SR-MPLS still requires coordinating a globally consistent segment routing global block (SRGB) to allocate SIDs, whereas SRv6 completely eliminates label coordination by using IPv6 address prefixes as SIDs and can summarize SIDs like regular IPv6 routes. In this research, we adopt and build analysis on the more advanced SRv6 that inherently offers stronger flexibility and scalability.



\begin{figure}[H]
    \centering
    \vspace{-5pt}
    \includegraphics[width=.485\textwidth]{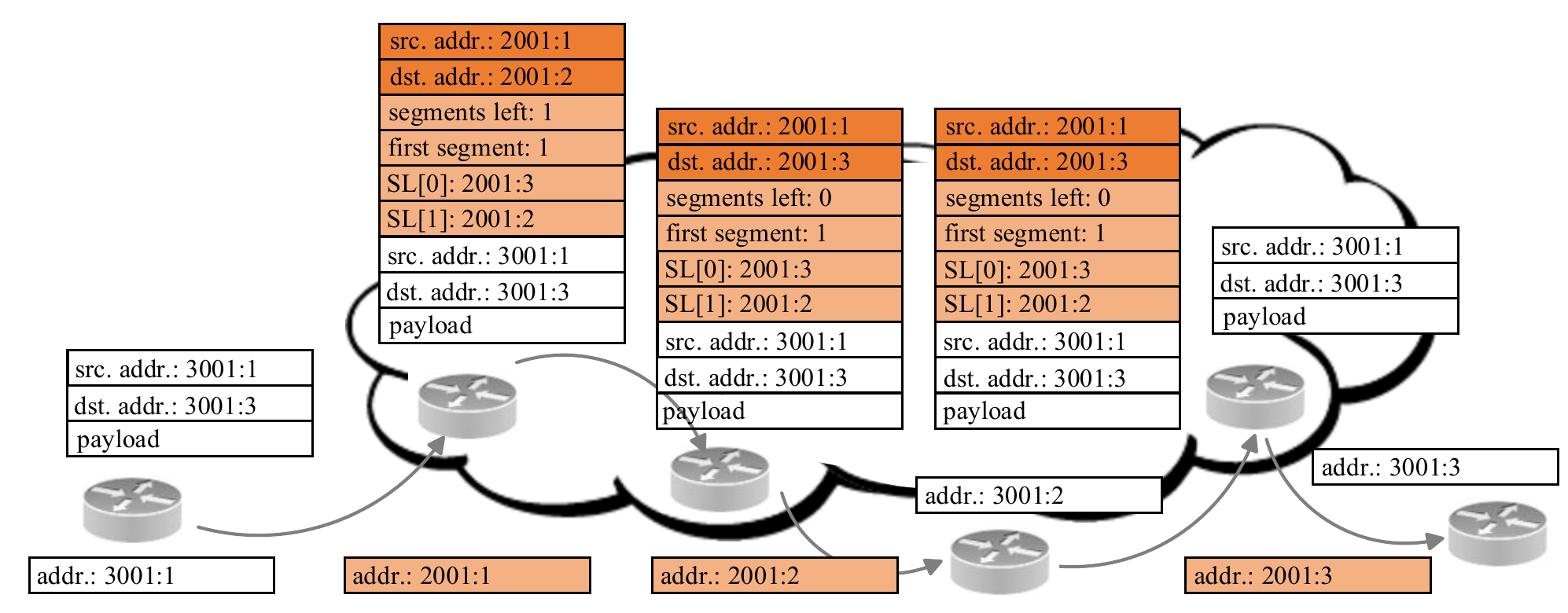}
    \vspace{-15pt}
    \caption{SRD 2001 consists of SRv6 nodes 2001:1-2001:3 and the IP domain consists of IP nodes 3001:1-3001:3. Each packet diagram denotes the packet header when emitted through the egress edge, in which the fields include the outer/encapsulated header (in dark orange), the SRH (in orange), and the payload (including inner header, etc.), resp., with the active segment always appearing in the outer header.}
    \vspace{-5pt}
    \label{fig:SRH}
\end{figure}

Since the Internet is expanding, segment routing domains (SRDs) \cite{ToN17} have been introduced primarily for scalability, administrative control, and security isolation. Informally, an SRD consists of a number of SR-capable nodes, and the routing details within the SRD can be hidden from the outside. To this end, for instance, SRDs use binding SIDs (BSIDs) that enable boundary nodes to handle intra-domain routing, expanding the BSID into complete path that can adapts to internal changes. Hybrid SR/IP networks exist in the gradual hardware migration process, in which SR-capable nodes interoperate with legacy IP nodes, but have significant implications for SRD topologies: SR-MPLS requires each SRD form a connected subnet since IP nodes will discard packets, while SRv6 lifts this restriction by allowing IP nodes to simply ignore the SRH and forward packets using legacy IGP rules \cite{ToN20}. Consequently, SR-capable nodes can be deployed dispersedly with SRv6 \cite{ToN20}, although fragmented SRDs remain administratively undesirable \cite{sr}. Fig. \ref{fig:SRH} illustrates a typical path in a hybrid IPv6 network, with SR-capable nodes 2001:1–2001:3 within SRD 2001.
In an SR/IP hybrid network that support multi-SRD partitions, every SR node can program SIDs in the forwarding information base (FIB). In addition, the SRD boundary nodes (i.e., ingress/egress nodes) assume greater responsibility than the interior nodes, such as mapping traffic flows to SR policies and performing SRH removal and packet decapsulation \cite{mo1}. The FIB use ternary content-addressable memory (TCAM), a device particularly well-suited for rapid longest prefix match (LPM) lookups on destination addresses. Although TCAM enables fast SID lookups, it is relatively expensive, consumes significant power, and has limited capacity, all rendering space conservation a persistent engineering challenge. Furthermore, hardware has hard limits on the number of equivalent paths it can support. For example, some Extreme Networks switches support a maximum of 64 ECMP gateways \cite{1-64-2}, and NVIDIA's ASIC supports a large number of ECMP next hops but the hash bucket count is only configurable as power-of-two terms from 64 to 4096 \cite{1-64-1}; these numerical ceilings directly determine the finest granularity of splitting ratios.

In this work, we consider combining together two simple schemes to alleviate TCAM limits: allowing same-destination flows to share the same FIB entry, a common practice adopted by commercial vendors like CISCO \cite{cisco}; also, employing the set of power-of-two terms to approximate the target splitting ratios, respecting the hardware precision limit while reducing TCAM overhead \cite{ecmp1}. We explicitly analyze the consequence of quantizing splitting ratios to integer multiples of power-of-two fractionals (such as 1/64) with a probability-theoretic argument using concentration inequalities to upper-bound the performance degradation, and conduct extensive experiment to validate our analysis. To the authors' best knowledge, there is currently no similar work that has established such a bound.

\section{System Model}

We assume the common setting during the SRv6 gradual deployment process. An SR/IP hybrid network $G=(\mathcal{V},\mathcal{E})$ consists of SR/IP nodes $\mathcal{V}$ and directed edges $\mathcal{E}$. The SR nodes support SRv6 and are decomposed into $K\geq1$ disjoint self-governing SRDs, while the IP nodes only support native IPv6. Given an SR policy with per-flow splitting ratios precomputed with arbitrary techniques such as linear programming (LP) to improve traffic engineering (TE), an ingress node $r$ can compute the source-agnostic splitting ratios locally, and steer all same-destination flows into an identical source-agnostic policy. For convenience, we list frequently encountered symbols and notations in Table. \ref{tab:frequently_used_notations}.

\vspace{-5pt}
\begin{table}[ht]
    \centering
    \caption{Frequently Used Notations}
    \label{tab:frequently_used_notations}
    \begin{tabular}{m{1.8cm} m{6.2cm}}
        \hline
        \vspace{1pt}
        \textbf{symbol} & \textbf{description} \\
        \hline
        \vspace{1pt}
        $\mathrm{SRD}(n)$ & SRD that SR node $n$ belongs to \\
        \vspace{1pt}
        $s,d,r$ & source node $s$, destination node $d$, boundary node $r$ \\
        \vspace{1pt}
        $e$ & edge $e=(e_1,e_2)$ \\
        \vspace{1pt}
        $t^{sd}$ & intensity of flow from $s$ to $d$\\
        \vspace{1pt}
        $t^{sd}_r$ & portion of $t^{sd}$ entering $\mathrm{SRD}(r)$ via ingress node $r$ \\
        \vspace{1pt}
         $t^{sd}_{re}$ & portion of $t^{sd}_{r}$ that traverses edge $e$ \\
        \vspace{1pt}
         $t^{d}_{r},t^{d}_{re}$ & $\sum_st^{sd}_r,\sum_st^{sd}_{re}$ \\
        \vspace{1pt}
        $\mathrm{path}^{d}_{r,n}$ & $n$-th path for flow $t^{d}_r$ until it reaches the next ingress node or $d$ \\
        \vspace{1pt}
        $\mathrm{path}^{sd}_{s,n}$ & $n$-th path for flow $t^{sd}_s$ (i.e., $t^{sd}$) until it reaches the next ingress node or $d$ \\
        \vspace{1pt}
        $t^{d}_{r,n},p^{d}_{r,n}$ & portion, fraction of flow $t^{d}_r$ that traverses $\mathrm{path}^{d}_{r,n}$ \\
        \vspace{1pt}
        $\hat{f}$ & r.v. corresponding to $f$ induced by quantization \\
        \hline
    \end{tabular} \\
    \vspace{1pt}
    By convention, intensity/portion means traffic and fraction means ratio.
\end{table}
\vspace{-5pt}

To achieve this, an ingress node $r$ associated with $\mathrm{SRD}(r)$ inspects the packet's destination, collects the corresponding flow intensity statistics, and computes the splitting ratios as
\begin{align}\label{eq:splitting_ratio}
    p^{d}_{r,n}=\frac{\sum_st^{sd}_{r,n}}{\sum_n\sum_st^{sd}_{r,n}},
\end{align}
where $(r,n)$ indexes distinct paths within $\mathrm{SRD}(r)$ (sorted lexicographically) that enters $\mathrm{SRD}(r)$ via ingress node $r$, and $t^{sd}_{r,n}$ denotes the portion of traffic intensity $t^{sd}$ (in, e.g., bits per second or packets per second) that enters $\mathrm{SRD}(r)$ via $r$ and is forwarded along path $(r,n)$. Clearly, source-agnostic splitting ratios can potentially reduce the number of required SR policies by up to a factor of $O(|\mathcal{V}|)$.

The quantization step should ensure all resulting $\hat{p}^d_{r,n}$ be an integer multiple of $1/64$, while satisfying the nonnegativity constraint $\hat{p}^d_{r,n}\geq0$ and normalization constraint $\sum_n\hat{p}^d_{r,n}=1$. The simple separate rounding and adjustment scheme introduces bias; the multinoulli sampling is unbiased, but cannot guarantee the rounding error $|\hat{p}^d_{r,n}-{p}^d_{r,n}|$ is tightly bounded and is therefore prone to high variance. The key idea is systemtic rounding (a.k.a. dependent rounding or systematic sampling), which we summarize in Lemma 1 and Alg. \ref{alg:rr-sum}.

\begin{algorithm}[htbp]
\caption{Systematic Rounding with Sum Constraint}
\label{alg:rr-sum}
\begin{algorithmic}[1]
\REQUIRE multinoulli distribution $p = ({p}^d_{r,1},\dots,{p}^d_{r,N})$ with ${p}^d_{r,n} \geq 0$ and $\sum_{n=1}^N {p}^d_{r,n} = 1$.
\ENSURE integers $(q_1,\dots,q_N)$ and $\hat{p}^d_{r,n} = q_n/64$ such that
         $\sum_{n=1}^{N} \hat{p}^d_{r,n}\! =\! 1$, $\mathbb{E}[\hat{p}^d_{r,n}] = {p}^d_{r,n}$, and $|\hat{p}^d_{r,n} - {p}^d_{r,n}|\! <\! 1/64$.
\STATE $x_n \gets 64\,{p}^d_{r,n}$ \textbf{for all} $n = 1,\dots,N$
\STATE $s_0 \gets 0$
\FOR{$n = 1$ \textbf{to} $N$}
    \STATE $s_n \gets s_{n-1} + x_n$
\ENDFOR
\STATE draw $U \sim \mathrm{Unif}[0,1)$
\FOR{$n = 1$ \textbf{to} $N$}
    \STATE $q_n \gets \lfloor s_n + U \rfloor - \lfloor s_{n-1} + U \rfloor$
    \STATE $\hat{p}^d_{r,n} \gets q_n / 64$
\ENDFOR
\RETURN $(\hat{p}^d_{r,1},\dots,\hat{p}^d_{r,N})$
\end{algorithmic}
\end{algorithm}

\noindent\textit{Lemma 1 (Systematic Rounding with a Sum Constraint)}. The systematic rounding algorithm produces $(\hat{p}^d_{r,1},\dots,\hat{p}^d_{r,N})$ that are nonnegative, normalized, unbiased,$\!$ and have bounded error.\\
\noindent\textit{Proof}. $s_n\geq s_{n-1}$ implies $q_n\geq0$, and
\begin{align*}
    \textstyle\sum_{n=1}^{N} q_n = \sum_{n=1}^{N}(\lfloor s_n + U\rfloor - \lfloor s_{n-1}+U\rfloor) = 64,
\end{align*}
which holds for all $U\!\in\![0,1)$ since the floors $\lfloor\cdot\rfloor$ telescope. For any real $t\geq0$ and $U\sim\mathrm{Unif}[0,1)$, it is easy to verify that $\mathbb{E}[\lfloor t+U\rfloor]=t$, and therefore $\mathbb{E}[q_n]=s_n-s_{n-1}=x_n$ by the linearity of expectations. Note that $q_n\in\{\lfloor x_n\rfloor, \lceil x_n\rceil\}$, then $|\hat{p}^d_{r,n} - {p}^d_{r,n}| = |q_n - x_n|/64 < 1/64$. $\hfill\square$ \\

Each ingress node $r$ can locally compute the source-agnostic splitting ratios with Eq. \eqref{eq:splitting_ratio} and quantize them with Alg. \ref{alg:rr-sum} for each destination address $d$. Once the source-agnostic policies have been established, ingress node $r$ looks up the destination IP address in its SR policy table and forwards the flows with corresponding $\hat{p}^{d}_{r,n}$.

\section{Degradation Bound Analysis}\label{sec:bsids}

Before formal analysis, we reiterate several practical conventions. First, in an SR/IP hybrid network $G$ with multiple SRDs, packets originating from an IP node cannot participate in SR, and packets originating from an SR node may have several ingress nodes (i.e., source/domain entry) and several egress nodes (i.e., destination/domain exit). Second, SR permits multiple traversals of the same SRD to improve TE performance by leveraging external paths \cite{ToN17}, as exemplified by IP node 3001:2 in Fig. \ref{fig:SRH}; however, policy nesting is prohibited to avoid routing loops and segment list depth (SLD) violations \cite{mo1}, and we assume default IP routing outside the current SRD before the packet is decapsulated at the egress node. This requirement limits cascading errors and guarantees performance, since quantization occurs at most $K$ times for any SR route in a $K$-SRD network. Third, while multiple traversals are allowed, we strictly forbid routing loops. Fourth, we adopt the systematic sampling approach in Alg. \ref{alg:rr-sum} to round the split ratios to integer multiple of 1/64 in compliance with the hardware precision limit.

Let $t^{sd}_{r}$ be the portion of $t^{sd}$ (in traffic intensity) that enters $\mathrm{SRD}(r)$ via ingress node $r$, and let $t^{sd}_{re}$ be the portion of $t^{sd}_{r}$ that traverses edge $e$. We can have $t^{sd}_{re}$ well defined only when the splitting ratio $t^{sd}_{re}/t^{sd}_r$ is computed at ingress node $r$, for which we can write $e\in\mathrm{SRD}(r)$; in this sense, any edge $e$ can belong to multiple SRDs for flow $t^{sd}$. Given traffic $t^{sd}$ and ingress node $r$, any edge $e$ can carry traffic from two types of sources: traffic originating inside $\mathrm{SRD}(r)$ (i.e., $r=s$) and traffic originating outside $\mathrm{SRD}(r)$ (i.e., $r\not=s$). For any portion or fraction $f$, let $\hat{f}$ denote its version induced by quantization.
\begin{figure}[H]
	\centering
	\vspace{-5pt}
	\includegraphics[width=.485\textwidth]{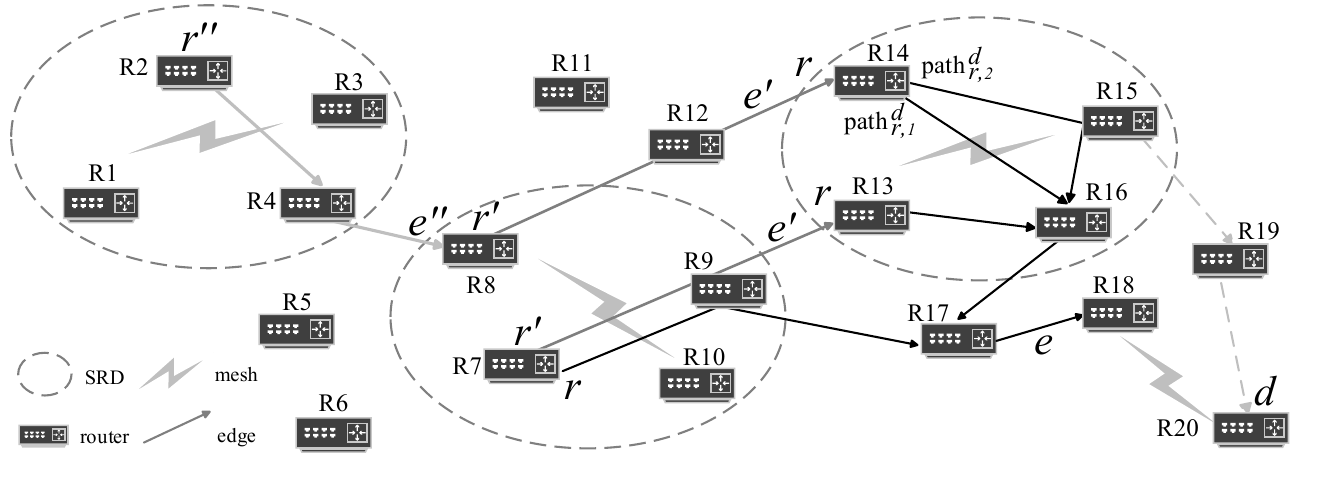}
	\vspace{-20pt}
	\caption{An illustration of SRDs in the SR/IP hybrid network. The edge $e$ of our primary interest can be inside or outside the SRDs; however, all SR flows though $e$ to $d$ must have corresponding ingress nodes $r$ and possibly $r'$, etc.}
	\vspace{-5pt}
	\label{fig:SRD}
\end{figure}
For an ingress node $r$, we propose forward decomposition of $t^d_{re}$ in Eq. \eqref{eq:forward} and backward decomposition of $t^d_r$ in Eq. \eqref{eq:backward}, respectively, which hold for both the original and quantized versions and is illustrated in Fig. \ref{fig:SRD}.:
\begin{align}\label{eq:forward}
	t^d_{re}=\textstyle\sum_{n}\mathbb{I}[e\in\mathrm{path}^d_{r,n}]t^d_rp^d_{r,n},
\end{align}
\begin{align}\label{eq:backward}
	t^d_r=\textstyle\sum_{r',e':e_2'=r,e'\in\mathrm{SRD}(r')}t^d_{r'e'},
\end{align}
where we use $r',e'$ to mean ingress nodes and edges for all possible immediate precursor SRDs, respectively. We prove that the per-edge traffic remains unbiased in Lemma 2.\\

\noindent\textit{Lemma 2 (Unbiased Per-Edge Traffic from Quantization)}. For any edge $e$, the per-edge traffic $\hat{t}_e$ remains unbiased using the quantized $\hat{p}^{d}_{r,n}$.\\
\noindent\textit{Proof}. Using conditional expectation $\mathbb{E}[X]=\mathbb{E}_Y[\mathbb{E}_X[X|Y]]$, letting $Y$ be $\hat{t}^{d}_r$, and noting that $\hat{p}^d_{r,n}$ is independent from $\hat{t}^{d}_r$,
\begin{align*}
	\mathbb{E}[\hat{t}^d_{re}]=\textstyle\sum_{n}\mathbb{I}[e\in\mathrm{path}^d_{r,n}]\mathbb{E}[\hat{t}^d_r]p^d_{r,n}.
\end{align*}
Taking expectation on Eq. \eqref{eq:backward},
\begin{align*}
	\mathbb{E}[\hat{t}^d_r]=\textstyle\sum_{r',e':e_2'=r,e'\in\mathrm{SRD}(r')}\mathbb{E}[\hat{t}^d_{r'e'}].
\end{align*}
By recursively unrolling the sums with Eqs. \eqref{eq:forward} and \eqref{eq:backward}, we know $\mathbb{E}[\hat{t}^d_{re}]=t^d_{re}$. Then $\mathbb{E}[\hat{t}_{e}]=\mathbb{E}[\sum_{d,r}\hat{t}^d_{re}]=\sum_{d,r}t^d_{re}=t_{e}$ for any $e$. $\hfill\square$\\

Before characterizing the degradation of performance, we propose Lemma 3 as a useful fact, and introduce Bernstein's inequality to yield pretty tight tail bounds with variance.\\

\noindent\textit{Lemma 3 (Variance Upper Bound for Partial Sum of Probabilities)}. For $\{\hat{p}^d_{r,n}\}$ generated by Alg. \ref{alg:rr-sum} and any $\mathcal{S}\!\subset\!\{1,\ldots,N\}$, $\text{Var}(\sum_{n\in\mathcal{S}}\hat{p}^d_{r,n})$ is upper bounded by $(1/4)(1/64)^2$.\\
\noindent\textit{Proof}. For a contiguous block $\mathcal{S}=\{N_1,\ldots,N_2\}$, telescoping using lines 6-9 of Alg. \ref{alg:rr-sum} gives
\begin{align*}
\textstyle\sum_{n\in\mathcal{S}}\hat{p}^d_{r,n}=\big(\lfloor\sum_{n=1}^{N_2}x_{n}\!+\!U\rfloor\!-\!\lfloor\sum_{n=1}^{N_1-1}\!x_{n}\!+\!U\rfloor\big)/64.
\end{align*}
Then $\text{Var}(\sum_{n\in\mathcal{S}}\hat{p}^d_{r,n})$ is equivalent to that of quantizing the single probability $\sum_{n=N_1}^{N_2}\!{p}^d_{r,n}$, given by
\begin{align*}
\textstyle\text{Var}(\sum_{n=N_1}^{N_2}\hat{p}^d_{r,n})=\{\sum_{n=N_1}^{N_2}\!{p}^d_{r,n}\}\big(1\!-\!\{\sum_{n=N_1}^{N_2}\!{p}^d_{r,n}\}\big)/{(64)^2},
\end{align*}
where $\{\cdot\}$ denotes the fractional part, and the upper bound is $(1/4)(1/64)^2$. For general $\mathcal{S}$ consisting of $R$ contiguous blocks, the variance cannot exceed $R^2(1/4)(1/64)^2$. Since $\mathcal{S}$ consists of at most $\lceil N/2\rceil$ contiguous blocks, we obtain the upper bound $\text{Var}(\sum_{n\in\mathcal{S}}\hat{p}^d_{r,n})\leq(\lceil N/2\rceil^2/4)(1/64)^2$. In addition, Popoviciu's inequality has established that
$\text{Var}[X]\leq{(b-a)^2}/{4}$ for random variable $X\!\in\![a,b]$. Then, in summary we have $\text{Var}(\sum_{n\in\mathcal{S}}\hat{p}^d_{r,n})\!\leq\!\min\{(\lceil\!N/2\rceil^2\!/4)(1/64)^2,1/4\}$. $\hfill\square$\\
\noindent\textit{Remark}. In contrast, the mean of 64 multinoulli sampling yields a variance of $\sum_{n\in\mathcal{S}}\hat{p}^d_{r,n}(1-\sum_{n\in\mathcal{S}}\hat{p}^d_{r,n})/64\!\leq\!1/256$, which exceeds the bound from Lemma 3 whenever $N\leq16$. In practice, however, we select only a few $\mathrm{path}^d_{r,n}$, a number far smaller than this threshold.\\

Then we prove that the per-edge traffic deviates from the mean with exponentially small probabilities in Lemma 4.\\ 

\noindent\textit{Lemma 4 (Exponentially Small Probability of Deviation in Per-Edge Traffic from Quantization)}. For any edge $e$, the per-edge traffic $\hat{t}_e$ deviates from $t_e$ by more than $t>0$ with exponentially small probability. That is,
\begin{align*}
\mathrm{Pr}\big( \big| \textstyle\!\sum_{d}(\hat{t}^d_e\! -\! {t}^d_e) \big|\! \geq\! t \big)\!
\leq\! 2\!\displaystyle \exp\!\left( \!-\frac{t^2}{2A\!\sum_{d}({t}^d)^2\! +\! {2} t \max_d{t}^d/{3}}\! \right),
\end{align*}
where $\Delta=\min\{(\lceil N/2\rceil^2/4)(1/64)^2,1/4\}$ from Lemma 3, $A\!=\!(1\!+\!\Delta)^K\!-\!1$, $d$ ranges over all possible destinations for which $t^d_e>0$, and $t^d=\sum_st^{sd}$.\\
\noindent\textit{Proof}. For any edge $e$, we attempt to obtain an upper bound on $\text{Var}(\hat{t}_e)$. To this end, we decompose $\text{Var}(\hat{t}^d_r)$ with conditional variance $\text{Var}(X)=\mathbb{E}_{Y}[\text{Var}_X(X|Y)]+\text{Var}_{Y}(\mathbb{E}_X[X|Y])$ and identifying $Y$ with $\{\hat{t}^d_{r'}\}$,
\begin{align}
	\text{Var}(\hat{t}^d_r)\stackrel{a}{=}&\ \mathbb{E}_Y[\text{Var}(\textstyle\sum_{r',e':e_2'=r,e'\in\mathrm{SRD}(r')}\hat{t}^d_{r'e'}|Y)]\notag\\
	& +\text{Var}_Y(\mathbb{E}_Y[\textstyle\sum_{r',e':e_2'=r,e'\in\mathrm{SRD}(r')}\hat{t}^d_{r'e'}|Y])\notag\\
	\stackrel{b}{=}& \textstyle\sum_{r'}\!\mathbb{E}_Y[\text{Var}(\textstyle\sum_{e':e_2'=r,e'\in\mathrm{SRD}(r')}\!\hat{t}^d_{r'e'}|Y)]\notag\\
	& +\textstyle\sum_{r'}\!\text{Var}_Y(\mathbb{E}_Y[\sum_{e':e_2'=r,e'\in\mathrm{SRD}(r')}\!\hat{t}^d_{r'e'}|Y]),\label{eq:decom1}
\end{align}
where step (a) follows from Eq. \eqref{eq:backward} and step (b) follows from independence among all $r'$ given $Y$.
Consider the summand in the first RHS term in Eq. \eqref{eq:decom1},
\begin{align}\label{eq:decom2}
	&\ \textstyle\mathbb{E}_Y[\text{Var}(\textstyle\sum_{e':e_2'=r,e'\in\mathrm{SRD}(r')}\hat{t}^d_{r'e'}|Y)]\notag\\
	\stackrel{c}{=}&\ \! \textstyle\mathbb{E}_Y[\text{Var}(\sum_{e':e_2'=r,e'\in\mathrm{SRD}(r')}\sum_{n}\mathbb{I}[e'\!\in\!\mathrm{path}^d_{r',n}]\hat{t}^d_{r'}\hat{p}^d_{r',n}|Y)]\notag\\
	\stackrel{d}{=}&\ \! \textstyle\mathbb{E}_Y[\text{Var}(\sum_{n}\mathbb{I}[e_2'\!=\!r,e'\!\in\!\mathrm{SRD}(r'),e'\!\in\!\mathrm{path}^d_{r',n}]\hat{t}^d_{r'}\hat{p}^d_{r',n}|Y)]\notag\\
	\stackrel{e}{\leq}&\ \!\textstyle\Delta\mathbb{E}[(\hat{t}^d_{r'})^2],
\end{align}
where step (c) follows by inserting in Eq. \eqref{eq:forward}, step (d) reduces to single sum because given $r'$, any $e'\in\mathrm{path}^d_{r',n}$ can hold for most one $n$, and step (e) follows from Lemma 3. Consider the summand in the second RHS term in Eq. \eqref{eq:decom1},
\begin{align}\label{eq:decom3}
&\ \!\textstyle\text{Var}_Y(\mathbb{E}[\textstyle\sum_{e':e_2'=r,e'\in\mathrm{SRD}(r')}\hat{t}^d_{r'e'}|Y)]\notag\\
=&\ \!\textstyle\text{Var}_Y(\mathbb{E}[\sum_{e':e_2'\!=\!r,e'\in\mathrm{SRD}(r')}\sum_{n}\mathbb{I}[e'\!\in\!\mathrm{path}^d_{r',n}]\hat{t}^d_{r'}\hat{p}^d_{r',n}|Y)]\notag\\
=&\ \!\textstyle\text{Var}_Y(\mathbb{E}[\sum_{n}\mathbb{I}[e_2'\!=\!r,e'\!\in\!\mathrm{SRD}(r'),e'\!\in\!\mathrm{path}^d_{r',n}]\hat{t}^d_{r'}\hat{p}^d_{r',n}|Y)]\notag\\
=&\ \!\textstyle\text{Var}_Y(\sum_{n}\mathbb{I}[e_2'=r,e'\!\in\!\mathrm{SRD}(r'),e'\!\in\!\mathrm{path}^d_{r',n}]\hat{t}^d_{r'}p^d_{r',n})]\notag\\
\leq&\ \!\textstyle\text{Var}[(\hat{t}^d_{r'})^2],
\end{align}
where the last step uses $0\leq\sum_{n}\mathbb{I}[e_2'=r,e'\in\mathrm{SRD}(r'),e'\in\mathrm{path}^d_{r',n}]p^d_{r',n}\leq1$. Since $\mathbb{E}[(\hat{t}^d_{r})^2]=(\mathbb{E}[\hat{t}^d_{r}])^2+\text{Var}(\hat{t}^d_{r})$ and $\mathbb{E}[\hat{t}^d_{r}]=t^d_{r}$, recursively we decompose the term $\text{Var}(\hat{t}^d_r)$ by inserting Eqs. \eqref{eq:decom2} and \eqref{eq:decom3} to Eq. \eqref{eq:decom1},
\begin{align}\label{eq:decom4}
	\text{Var}(\hat{t}^d_r)\leq\textstyle\sum_{r'\in p(r)}(1\!+\!\Delta)\text{Var}(\hat{t}^d_{r'})\!+\!\sum_{r'\in p(r)}\Delta({t}^d_{r'})^2.
\end{align}
For each $r'$, similarly we can define $r''$ to have
\begin{align}\label{eq:decom5}
\text{Var}(\hat{t}^d_{r'})\!\leq\!\textstyle\sum_{r''\!\in\!p(r')}\!(1\!+\!\Delta)\text{Var}(\hat{t}^d_{r''})\!+\!\sum_{r''\!\in\!p(r')}\!\Delta({t}^d_{r''})^2,
\end{align}
Plugging Eq. \eqref{eq:decom5} to Eq. \eqref{eq:decom4}, we have
\begin{align}\label{eq:decom6}
	\text{Var}(\hat{t}^d_r)\!\leq\!&\textstyle\sum_{r'\in p(r)}(1\!+\!\Delta)\sum_{r''\in p(r')} (1\!+\!\Delta)\text{Var}(\hat{t}^d_{r''})\notag\\
	+&\textstyle\sum_{r'\!\in\!p(r)}\!(1\!+\!\Delta)\!\sum_{r''\!\in\!p(r')}\!\Delta({t}^d_{r''}\!)^2\!+\!\textstyle\sum_{r'\!\in\!p(r)}\!\Delta({t}^d_{r'}\!)^2\notag\\
	=&\textstyle\sum_{r''\in pp(r)}(1\!+\!\Delta)^2\text{Var}(\hat{t}^d_{r''})\notag\\
	+&\textstyle\sum_{r''\in pp(r)}\!\Delta (1\!+\!\Delta)({t}^d_{r''})^2+\!\textstyle\sum_{r'\in p(r)}\!\Delta({t}^d_{r'})^2,
\end{align}
where $p(r)$ denotes all possible ingress nodes in the immediate precursor SRD before $r$, and clearly the double sum $\sum_{r'\in p(r)}\sum_{r''\in p(r')}$ collapses into the single sum $\sum_{r''\in pp(r)}$. We can fully unroll the recursion back to source SRDs as
\begin{align}\label{eq:decom7}
	\text{Var}(\hat{t}^d_r)\leq&\textstyle\sum_{r^{K-1}\in p^{K-1}(r)}(1\!+\!\Delta)^{K-1}\text{Var}(\hat{t}^d_{r^{K-1}})\notag\\
	+&\!\textstyle\sum_{r^{K-1}\in p^{K-1}(r)}\Delta(1\!+\!\Delta)^{K-2}({t}^d_{r^{K-1}})^2\notag\\
	+&\!\textstyle\sum_{r''\!\in\!pp(r)}\!\Delta(1\!+\!\Delta)({t}^d_{r''})^2\!+\!\textstyle\sum_{r'\!\in\!p(r)}\!\Delta({t}^d_{r'})^2,
\end{align}
where the superscript ${K\!-\!1}$ denotes $(K-1)$-fold apostrophes. By conditioning on $\{\hat{t}^d_{r}\}$, we can verify
\begin{align}\label{eq:var_dre}
	\textstyle\text{Var}(\hat{t}^d_{e})\!=\!\text{Var}(\sum_r\hat{t}^d_{re})\!\leq\!\sum_r\!\big(\textstyle(1\!+\!\Delta)\text{Var}(\hat{t}^d_{r})\!+\!\Delta({t}^d_{r})^2\big).
\end{align}
Since ${t}^d_{r^{K-1}}$ is the sum of flows to $d$ whose ingress node in the first SRD is $r^{K-1}$, $r^{K-1}$ must be the source node $s$. Hence ${t}^d_{r^{K-1}}\!=\!t^{sd}$, which remains constant and unaffected by quantization, so $\text{Var}(\hat{t}^d_{r^{K-1}})\!=\!0$. Inserting Eq. \eqref{eq:decom7} to Eq. \eqref{eq:var_dre},
\begin{align}\label{eq:decom8}
	\textstyle\text{Var}(\hat{t}^d_{e})\leq&\textstyle\sum_{r^{K-1}\in p^{K-1}(r)}\Delta (1\!+\!\Delta)^{K-1}({t}^d_{r^{K-1}})^2+\ldots\notag\\
	&+\!\textstyle\sum_{r'\in p(r)}\Delta (1\!+\!\Delta)({t}^d_{r'})^2+\!\textstyle\sum_{r}\Delta({t}^d_{r})^2.
\end{align}
For given edge $e$, $t^{d}_{r}$ denotes all flows to $d$ that traverse ingress node $r$ and $e\!\in\!\mathrm{SRD}(r)$, so $\sum_rt^{d}_{r}\!\leq\!t^d$.

Our analysis relies on the fact that for any $1\!\leq\!k\!\leq\!K\!-\!1$, $\sum_r\sum_{r^k\in p^k(r)}t^{d}_{r^k}\!\leq\!\sum_rt^{d}_{r}$. To understand this, consider all $r'\!\in\!p(r)$. The sum $\sum_r\sum_{r'\in p(r)}t^{d}_{r'}$ means all flows to $d$ that traverse ingress node $r'$ immediately before ingress node $r$. The equality holds when every such flow has $r'$ preceding $r$, while the strict inequality holds when some of these flows do not contain $r'$, i.e., when they originate at $r$. Then we directly obtain $\text{Var}(\hat{t}^d_{e})\leq\left((1+\Delta)^K-1\right)(t^d)^2$ from Eq. \eqref{eq:decom8}. With the same argument, we also have for any $1\!\leq\!k\!\leq\!K\!-\!2$, $\sum_r\sum_{r^{k+1}\in p^{k+1}(r)}t^{d}_{r^{k+1}}\!\leq\!\sum_r\sum_{r^{k}\in p^{k}(r)}t^{d}_{r^{k}}$, a relation we do not use at this time. In Bernstein's inequality (see Appendix A), identify $n$ with $d$, $N$ with the set of $d$ for which $t^d_e>0$, and $X_n$ with $\hat{t}^d_{e}-t^d_{e}$, note that all $\hat{t}^d_e$ are independently quantized for different $d$, $\mathbb{E}[\hat{t}^d_{e}-t^d_{e}]=0$ by Lemma 2, and $|\hat{t}^d_{e}-t^d_{e}|\leq t^d$, and we obtain the result. $\hfill\square$

\section{Simulation Results}

We run experiments on the \texttt{ta2} instance, which is the largest standard benchmark in the survivable network design library (SNDlib) \cite{sndlib} and contains 65 nodes, 108 links, and 1869 demands. Its size and dense demand matrix make it a suitable testbed for studying how routing and quantization choices affect per-link traffic at scale. We vary two hyperparameters: the number of SRDs, $K \in \{5, 10\}$, which controls the topological granularity of the SRD partition, and the number of routes per demand, $N \in \{4, 8\}$, which influences the quantization error by spreading a demand over multiple paths. The quantization grid is fixed at $1/64$, as hard-coded in the implementation, so that the effect of $K$ and $N$ can be isolated.

The experimental results are shown in the plots. The $x$-axis is the integer index of each network edge, with each integer tick correspond to one physical edge of the \texttt{ta2} network. The index is not a physical quantity but a fixed enumeration of edges, so that the two sampling series and the Bernstein markers for the same edge land on the same horizontal position. The $y$-axis is a percentage. For the two sampling series, it is the absolute difference between the quantized and unquantized traffic intensity, expressed as a percentage of the unquantized intensity; for the three Bernstein series, it is the corresponding threshold $t$ divided by the unquantized traffic intensity, again as a percentage. Because both quantities share the same denominator, they are directly comparable on a single $y$-axis. The three Bernstein series correspond to probability values of $0.05$, $0.1$, and $0.5$, respectively.

\begin{figure}[t]
	\centering
	\vspace{-3pt}
	\includegraphics[width=.485\textwidth]{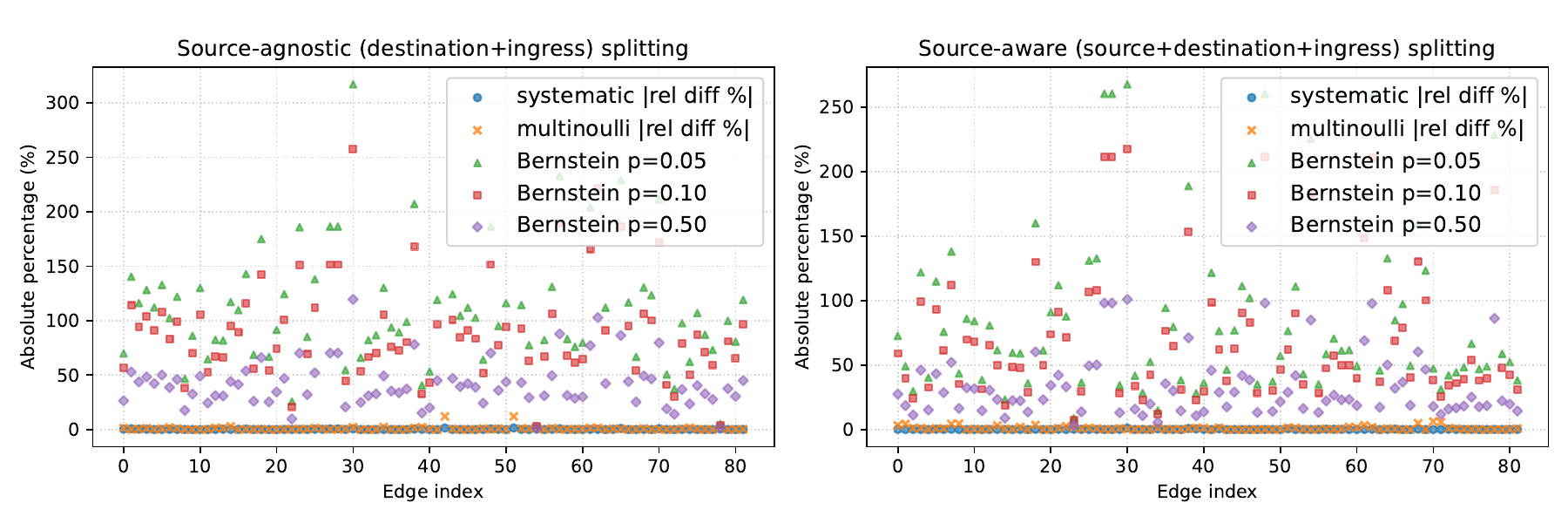}
	\vspace{-3pt}
	\includegraphics[width=.485\textwidth]{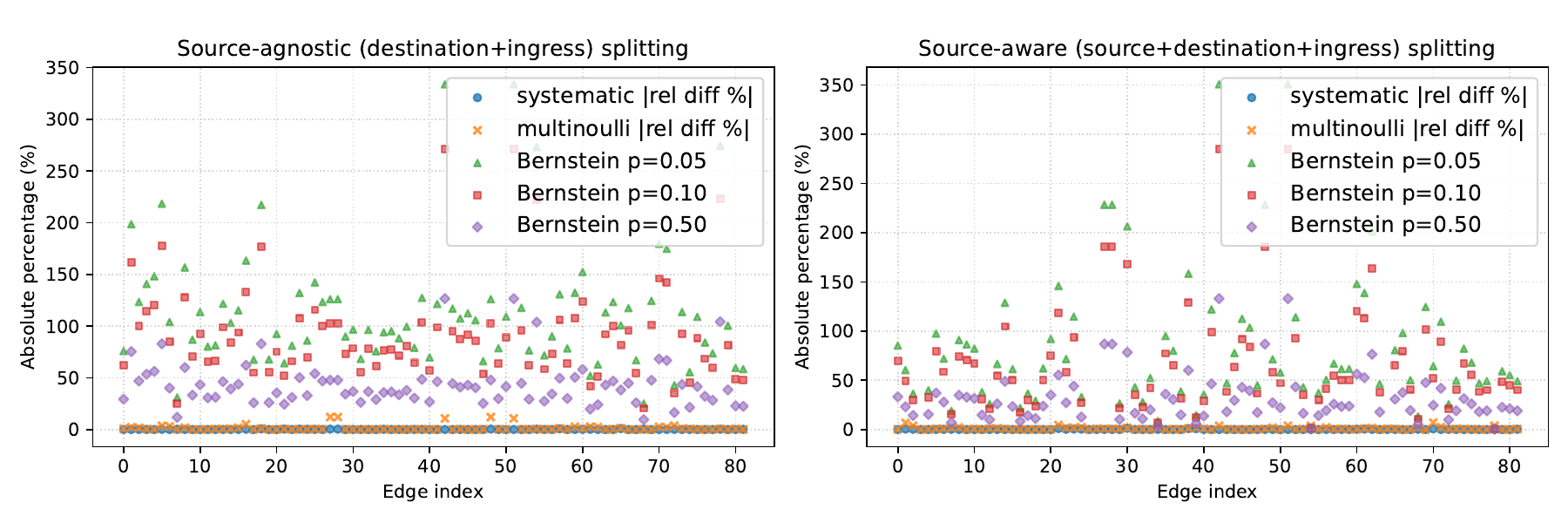}
	\vspace{-3pt}
	\includegraphics[width=.485\textwidth]{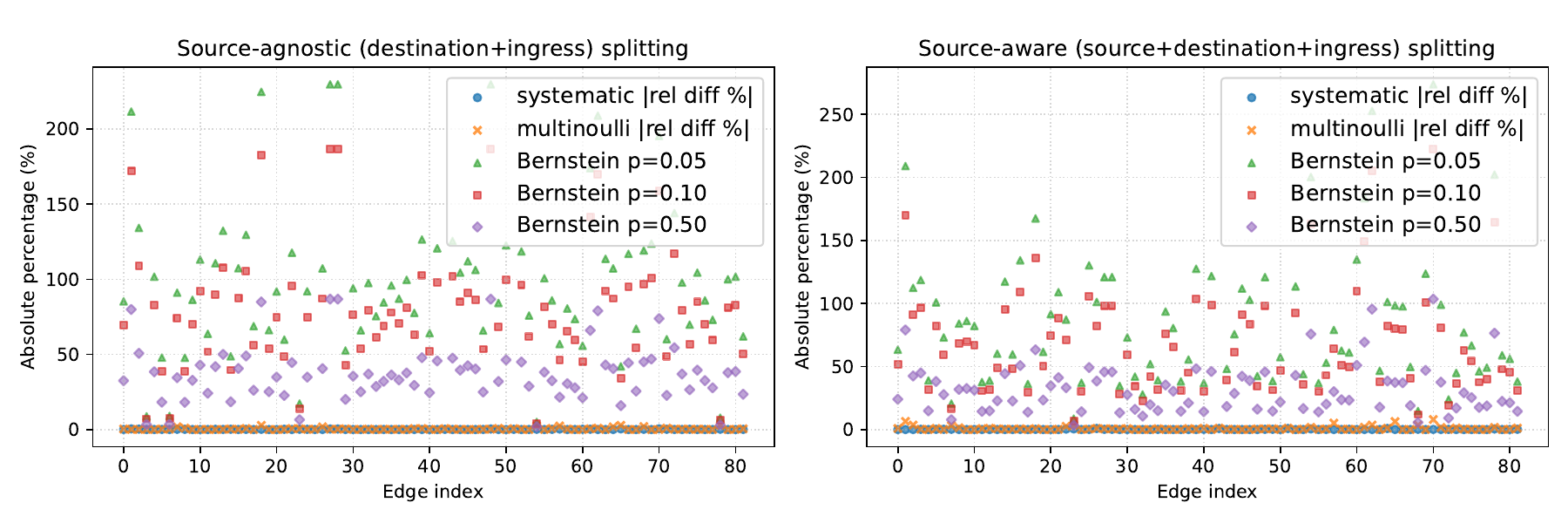}
	\vspace{-3pt}
	\includegraphics[width=.485\textwidth]{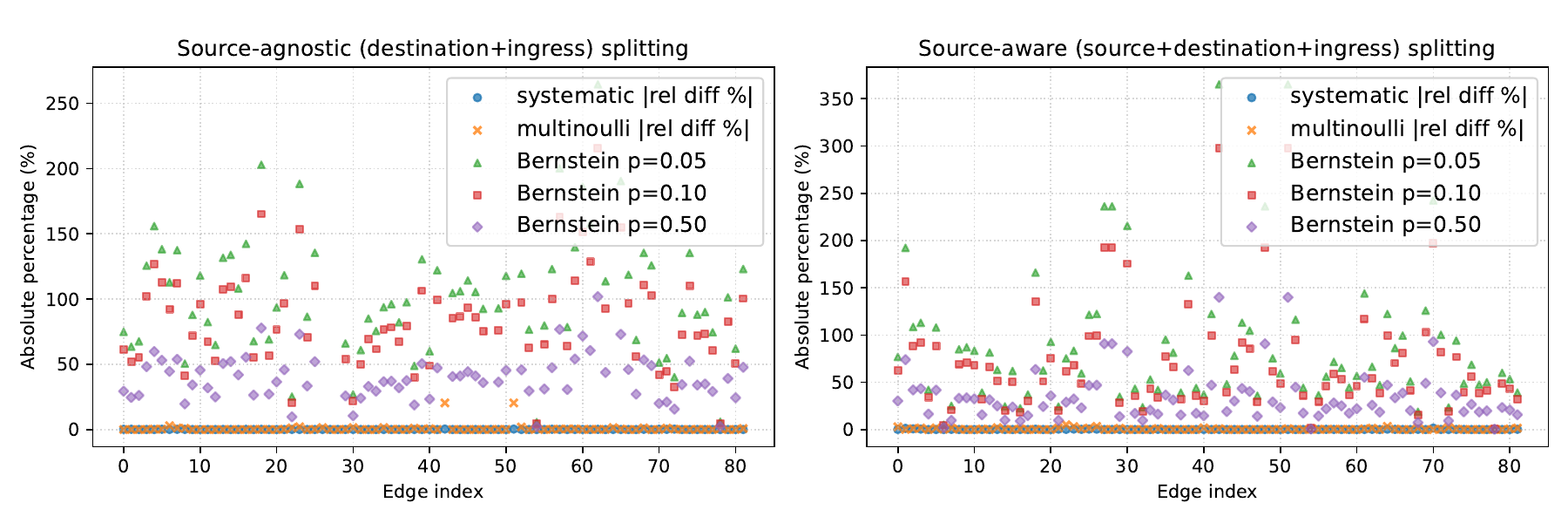}
	\vspace{-15pt}
	\caption{Relative error in percentage v.s. edge index for combinations of $K,N$ vertically and systematic/multinoulli sampling horizontally.}
	\vspace{-15pt}
\end{figure}

Comparing the plots, we can draw these conclusions.
\begin{itemize}
	\item First, source-aware splitting ratios yield a tighter Bernstein bound. This follows by replacing, in the summand of Bernstein's inequality, the aggregate over all sources $s$ for a fixed destination $d$ with the corresponding sum over specific $(s,d)$ pairs. The gain in tightness, however, is obtained at the expense of the principal advantage of source-agnostic splitting, which requires fewer splitting ratios to be cached in TCAM.
	\item Second, systematic quantization attains a more favorable error distribution than multinoulli quantization, the latter exhibiting larger variance. Multinoulli sampling draws a fixed number of independent samples, and no sharp bound is available on the absolute difference between the true and the quantized splitting ratios; this absence of a sharp bound can induce substantial variance.
	\item Third, almost all data points fall within the range associated with the $0.5$-probability bound, which suggests that the bound is not tight. Were the bound close to tight, approximately half of the points would lie within the $0.5$-probability range and the remainder outside it. The observed concentration therefore indicates that the Bernstein thresholds are conservative for this instance, at least under the hyperparameter settings considered here.
\end{itemize}

\section{Conclusion and Future Work}\label{sec:conclusion}

In this work, we analyze the consequence of SR splitting ratio quantization on the TCAM to respect the practical hardware precision constraints, while addressing the capacity limit and energy consumption challenges. Since SR's application scope is expanding from the Internet and data centers to satellite-terrestrial integrated networks and power grid communication networks, the methodology and results in this research work will have broader audience and impact. With the development of high-dimensional probabilistic inequality, the gap between the upper bound and empirical results can be further decreased.

\section*{Acknowledgments}

This research has been supported in part by the Singapore Ministry of Education Academic Research Fund Tier 2 MOE-T2EP20224-0009 and NTU startup.

\section*{Appendix A: Bernstein's Inequality}

\noindent\textit{Theorem 1 (Bernstein's Inequality)}. For independent random variables $\{X_n\}_{n=1}^N$ with $\mathbb{E}[X_n] = 0$ and $|X_n| \leq M$ a.s. for all $n$, let $\sigma^2 = \sum_{n=1}^{N} \text{Var}(X_n)$, and then for any $t > 0$,
\begin{equation*}
\mathrm{Pr}\left( \big| \textstyle\sum_{n=1}^{N} X_n \big| \geq t \right)
\leq 2\displaystyle \exp\left( -\frac{t^2}{2\sigma^2 + {2} M t/{3}} \right).
\end{equation*}

\vspace{12pt}

\end{document}